\documentclass[
reprint,
superscriptaddress,
amsmath,amssymb,
aps,
prl,
]{revtex4-1}

\usepackage{graphicx}
\usepackage{dcolumn}
\usepackage{upgreek}
\usepackage{mathrsfs}
\usepackage{setspace}
\usepackage{bm}
\usepackage[breaklinks=true,colorlinks=true,linkcolor=blue,urlcolor=blue,citecolor=blue]{hyperref}
\allowdisplaybreaks[4]
\begin{document}
	
	\preprint{APS/123-QED}
	
	\title{Deterministic Photon Sorting in Waveguide QED Systems}

	\author{Fan Yang}
	\affiliation{Center for Complex Quantum Systems, Department of Physics and Astronomy, Aarhus University, DK-8000 Aarhus C, Denmark}
	
	\author{Mads M. Lund}
	\affiliation{Center for Complex Quantum Systems, Department of Physics and Astronomy, Aarhus University, DK-8000 Aarhus C, Denmark}
	
	\author{Thomas Pohl}
	\affiliation{Center for Complex Quantum Systems, Department of Physics and Astronomy, Aarhus University, DK-8000 Aarhus C, Denmark}
	
	\author{Peter Lodahl}
	\affiliation{Center for Hybrid Quantum Networks (Hy-Q), Niels Bohr Institute, University of Copenhagen, Blegdamsvej 17, DK-2100 Copenhagen, Denmark}
	
	
	\author{Klaus M{\o}lmer}
	\email{moelmer@phys.au.dk}
	\affiliation{Center for Complex Quantum Systems, Department of Physics and Astronomy, Aarhus University, DK-8000 Aarhus C, Denmark}
	\affiliation{Aarhus Institute of Advanced Studies, Aarhus University, DK-8000 Aarhus C, Denmark}

	
	\begin{abstract}

		Sorting quantum fields into different modes according to their Fock-space quantum numbers is a highly desirable quantum operation. In this Letter, we show that a pair of two-level emitters, chirally coupled to a waveguide, may scatter single- and two-photon components of an input pulse into orthogonal temporal modes with a fidelity $\gtrsim 0.9997$. We develop a general theory to characterize and optimize this process and observe an interesting dynamics in the two-photon scattering regime: while the first emitter gives rise to a complex multimode field, the second emitter recombines the field amplitudes and the net two-photon scattering induces a self-time reversal of the pulse mode. The presented scheme can be employed to construct logic elements for propagating photons, such as a deterministic nonlinear-sign gate with a fidelity $\gtrsim 0.9995$.

	\end{abstract}
	
	\maketitle
	
	Strong nonlinearity at the few-photon level is key to all-optical quantum information processing (QIP) \cite{chang2014quantum}. In the last decade, quantum nonlinear optical phenomena have been demonstrated on various platforms \cite{chang2018colloquium}, including cavity/waveguide quantum electrodynamics (QED) setups \cite{arcari2014near}, atomic ensembles \cite{peyronel2012quantum}, and optomechanics \cite{aspelmeyer2014cavity}. Following these experimental achievements, the next step is to develop schemes that can utilize the acquired nonlinearity to perform high-fidelity QIP operations, such as single-photon transistors \cite{murray2017coherent}, single-photon subtractors \cite{yang2020atom}, and photonic logic gates \cite{brod2016passive,heuck2020controlled}. Among these quantum devices, a photon sorter which can separate single- and two-photon components of a single-mode input state into orthogonal output modes is particularly useful \cite{witthaut2012photon,ralph2015photon,bennett2016semiconductor,pick2021boosting}. 
	Ref.~\cite{ralph2015photon} thus proposed to employ the chiral coupling to a two-level emitter and scatter a single-mode input pulse into a field with the different number states occupying different photonic temporal modes \cite{brecht2015photon}.
	
\begin{figure}[b]
		\centering
		\includegraphics[width=\linewidth]{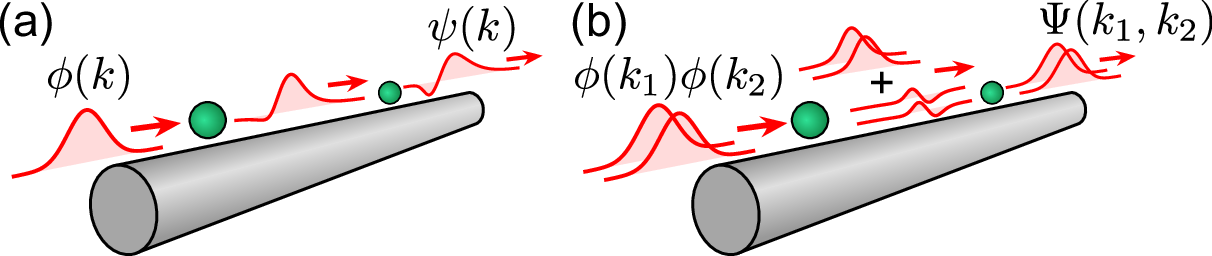}
		\caption{Sequential scattering of a single-photon pulse $\phi(k)$ and two-photon pulse $\Phi(k_1,k_2)=\phi(k_1)\phi(k_2)$ by a pair of chirally coupled two-level emitters. Panel (a) shows The linear dispersion of the single photon component; panel (b) shows the non-linear scattering of the two-photon state into multiple modes and back to a single mode output.}
		\label{fig:fig1}
	\end{figure}

	In this Letter, we establish a systematic approach to evaluate and optimize the performance of photon sorting in temporal-mode space. We find that
	the sorting by a single emitter, proposed in Ref.~\cite{ralph2015photon}, is hampered by a small but finite occupation of undesired modes, while, for a suitably optimized input pulse, the subsequent scattering on a second emitter restricts the one- and two-photon states to two orthogonal output modes with a very high fidelity (see Fig.~\ref{fig:fig1}). Our theoretical approach identifies a novel self-time-reversal mechanism of two-photon states by pairs of emitters, and we verify that our optimal photon sorting is robust against experimental imperfections. This makes it a  promising element in efficient Bell state analysis \cite{witthaut2012photon}, photonic controlled phase gates \cite{ralph2015photon}, as well as measurement-based quantum computing \cite{pick2021boosting}.
	
	{\it Model}.---We study a waveguide QED system, where an incident pulse interacts with $N_e$ two-level emitters in a unidirectional manner (see Fig.~\ref{fig:fig1} for $N_e=2$) \cite{tiecke2014nanophotonic,volz2014nonlinear,sollner2015deterministic,lodahl2017chiral}. We focus here on the scattering of a single-photon state $\hat{a}_{\phi}^\dagger|0\rangle$ and a two-photon state $(\hat{a}_{\phi}^\dagger)^2|0\rangle/{\sqrt{2}}$, where $\hat{a}_{\phi}^\dagger=\int dk \phi(k)\hat{a}^\dagger({k})$ creates a single photon in the input mode $\phi(k)$. We assume a unit propagation speed and $\hat{a}^\dagger(k)$ is the creation operator in  wave number (and frequency) space. The output states for the single- and two-photon input are respectively given by $\int dk \psi(k)\hat{a}^\dagger({k})|0\rangle$ and $\int dk_1 dk_2 \Psi(k_1,k_2)\hat{a}^\dagger({k_1})\hat{a}^\dagger({k_2})|0\rangle/\sqrt{2}$, where the single-photon pulse $\psi(k)=\mathcal{T}(k)\phi(k)$ is modified by a linear transmission coefficient $\mathcal{T}(k)$, while the two-photon wavefunction $\Psi(k_1,k_2)=\int dp_1 dp_2\mathcal{S}(k_1,k_2;p_1,p_2)\phi(p_1)\phi(p_2)$ is governed by the scattering matrix $\mathcal{S}$ \cite{shen2007strongly,fan2010input,mahmoodian2018strongly,lejeannic2021exper}. The cascaded feature of the chiral waveguide QED system allows us to obtain $\mathcal{T}(k)$ and $\mathcal{S}$ from the transmission coefficient $\mathcal{T}_0(k)$ and scattering matrix $\mathcal{S}_0$ solved for a single emitter with $\mathcal{T}=\mathcal{T}_0^{N_e}$ and $\mathcal{S}=\mathcal{S}_0^{N_e}$. For simplicity, we consider first the ideal case where photons are perfectly scattered into the guided mode of interest, i.e., $\psi(k)$ and $\Psi(k_1,k_2)$ have unit norm.
	
	
	\begin{figure}
		\centering
		\includegraphics[width=\linewidth]{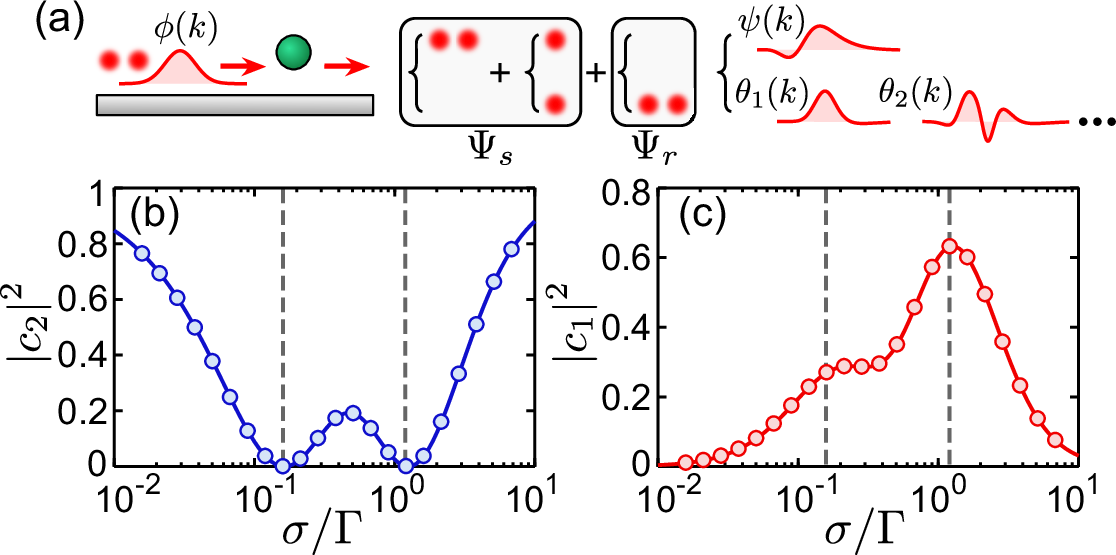}
		\caption{(a) Decomposition of the two-photon output wavefunction using single-photon pulse modes $\psi(k)$ and $\{\theta_n(k)\}$. (b) and (c) show the  probabilities $|c_2|^2$ and $|c_1|^2$ [c.f., Eq.~(\ref{eq:eq2})], as a function of the linewidth $\sigma$ of an input Lorentzian pulse. The circles and the solid lines are obtained by the quantum pulse method and Eqs.~(\ref{eq:eq4}) and (\ref{eq:eq5}), respectively.}
		\label{fig:fig2}
	\end{figure}
	To coherently split the single- and two-photon component of a superposed input state $\alpha\hat{a}_{\phi}^\dagger|0\rangle+\beta(\hat{a}_{\phi}^\dagger)^2|0\rangle/{\sqrt{2}}$, we require the two-photon output $\Psi(k_1,k_2)$ to occupy spatio-temporal modes which are orthogonal to the single-photon output $\psi(k)$ \cite{eckstein2011quantum,ansari2018tomography,ansari2018tailoring}. To explicitly quantify such a requirement, we need to expand the two-photon output state in terms of the single-photon wavefunction. First, the permutation symmetry of the bosonic wavefunction $\Psi(k_1,k_2)=\Psi(k_2,k_1)$ allows us to perform the Takagi factorization \cite{Paif2001quantum}
	\begin{equation}
	\Psi(k_1,k_2) =\textstyle \sum_{n}a_n f_n(k_1) f_n(k_2), \label{eq:eq1}
	\end{equation}
	where $\{f_n(k)\}$ forms a set of orthonormal basis functions. We then expand those basis functions on the single-photon output mode $\psi(k)$ and a normalized function ${\theta}_n(k)$, orthogonal to $\psi(k)$, i.e., $f_n(k)=\lambda_n\psi(k)+\mu_n{\theta}_n(k)$ With such a decomposition, the two-photon output state can be written as $\Psi(k_1,k_2)=\Psi_s(k_1,k_2)+\Psi_r(k_1,k_2)$, where the first part
	\begin{equation}
	\Psi_s = c_2\psi(k_1)\psi(k_2)+\frac{c_1}{\sqrt{2}}[\psi(k_1)\theta(k_2)+\theta(k_1)\psi(k_2)] \label{eq:eq2}
	\end{equation}
	corresponds to the state in which (i) both photons are populating the single-photon output mode $\psi(k)$ or (ii) only one of the photons is occupying $\psi(k)$ while the other is in an orthogonal mode $\theta(k)\propto \sum_n a_n\mu_n\lambda_n {\theta}_n(k)$ [see Fig.~\ref{fig:fig2}(a)].  The unwanted single and double occupation amplitudes $c_1$ and $c_2$ are determined by
	\begin{align}
	c_2 &= \int dk_1dk_2 [\psi(k_1)\psi(k_2)]^*\Psi(k_1,k_2),\label{eq:eq4}
	\\
	c_1\theta(k) &=  \sqrt{2}\int dk_1  \psi^*(k_1)\Psi(k_1,k)-\sqrt{2}c_2\psi(k).\label{eq:eq5}
	\end{align}
	The remaining two-photon wavefunction component
	\begin{equation}
	\Psi_r =\textstyle \sum_{n}a_n \mu_n^2 {\theta}_n(k_1) {\theta}_n(k_2) \label{eq:eq3}
	\end{equation}
	is the desired output as it contains no photons in the $\psi(k)$ mode [$\int dk_1 \psi^*(k_1)\Psi_r(k_1,k_2)=0$, see Fig.~\ref{fig:fig2}(a)]. A perfect photon sorter requires $c_1=c_2=0$ and, while it was shown in Ref.~\cite{ralph2015photon} that the condition $c_2=0$ can be satisfied in a single-emitter waveguide QED system by choosing a proper input pulse width, it is not clear whether the single excitation probability $|c_1|^2$ can be made simultaneously vanishing.
	
	To address this question, we first study the sorting performance of a single two-level emitter for a Lorentzian input pulse $\phi(k)\propto 1/(k^2+\sigma^2)$ with different spectral widths $\sigma$ in units of the coupling strength $\Gamma$ between the emitter and the guided mode. We extract the double and single excitation probabilities $|c_2|^2$ and $|c_1|^2$ by the relations established in Eqs.~(\ref{eq:eq4}) and (\ref{eq:eq5})  as well as by the input-output quantum pulse method \cite{kiilerich2019input,kiilerich2020quantum}. As shown in Figs.~\ref{fig:fig2}(b) and \ref{fig:fig2}(c), these two methods exhibit excellent agreement with each other, and they both identify perfect zeros of $|c_2(\sigma)|^2$ (dashed lines). They also show unfortunate maxima of $|c_1(\sigma)|^2$ for the same pulses, and this, eventually results in an imperfect sorting of Lorentzian input pulses with $|c_1|^2+|c_2|^2\geq 0.26$. 
	
	{\it Optimization protocol}.---The above results suggest that we need to establish a joint optimization protocol that takes both $c_1$ and $c_2$ into account in the search for the optimal input mode $\phi(k)$. To this end, we first define the sorting error to be $E = |c_1|^2+|c_2|^2$, and the corresponding sorting fidelity $\mathcal{F}=1-E$. With Eqs.~(\ref{eq:eq4}) and (\ref{eq:eq5}), the sorting error $E$ can be expressed as a functional of the input mode function $\phi(k)$, i.e., $E(\phi,\phi^*) = \int dp dp^\prime \phi^*(p) \mathcal{H}(p,p^\prime) \phi(p^\prime)$, where the Hermitian kernel $\mathcal{H}$ is given by
	\begin{equation}
	\mathcal{H}(p,p^\prime) = 2\int dk \mathcal{L}_1^*(k,p)\mathcal{L}_1(k,p^\prime)-\mathcal{L}_2^*(p) \mathcal{L}_2(p^\prime), \label{eq:eq6}
	\end{equation}
	with $\mathcal{L}_1(k,p)$ and $\mathcal{L}_2(p)$ defined as
	\begin{align}
	\mathcal{L}_1(k,p) &=\int dk_1dp_1 \mathcal{S}(k_1,k;p_1,p)\mathcal{T}^*(k_1)\phi^*(k_1)\phi(p_1), \nonumber \\
	\mathcal{L}_2(p) &=\int dk \mathcal{L}_1(k,p) \mathcal{T}^*(k)\phi^*(k). \nonumber
	\end{align}
	Now, the optimization problem amounts to finding the wavefunction that minimizes the error functional $E(\phi,\phi^*)$. The Hermitian kernel $\mathcal{H}$ is highly nonlinear in $\phi$, and this implies that the minimum  cannot be obtained by mere diagonalization or by a power iteration that work well for linear systems. In order to minimize $E(\phi,\phi^*)$, we therefore use the continuous steepest descent, which evolves a normalized gradient flow \cite{supply}
	\begin{equation}
	\partial_\tau \phi = -\frac{\delta E(\phi,\phi^*)}{\delta \phi^*}. \label{eq:eq7}
	\end{equation}
	In practice, we successively propagate Eq.~(\ref{eq:eq7}) for small time steps $\Delta \tau$ and renormalize $\phi$ to get a sequence that gradually diminishes $E(\phi,\phi^*)$, in the same spirit as the imaginary time evolution method leading to the ground state of an interacting Bose-Einstein condensate \cite{chiofalo2000ground}.
	
	\begin{figure}
		\centering
		\includegraphics[width=\linewidth]{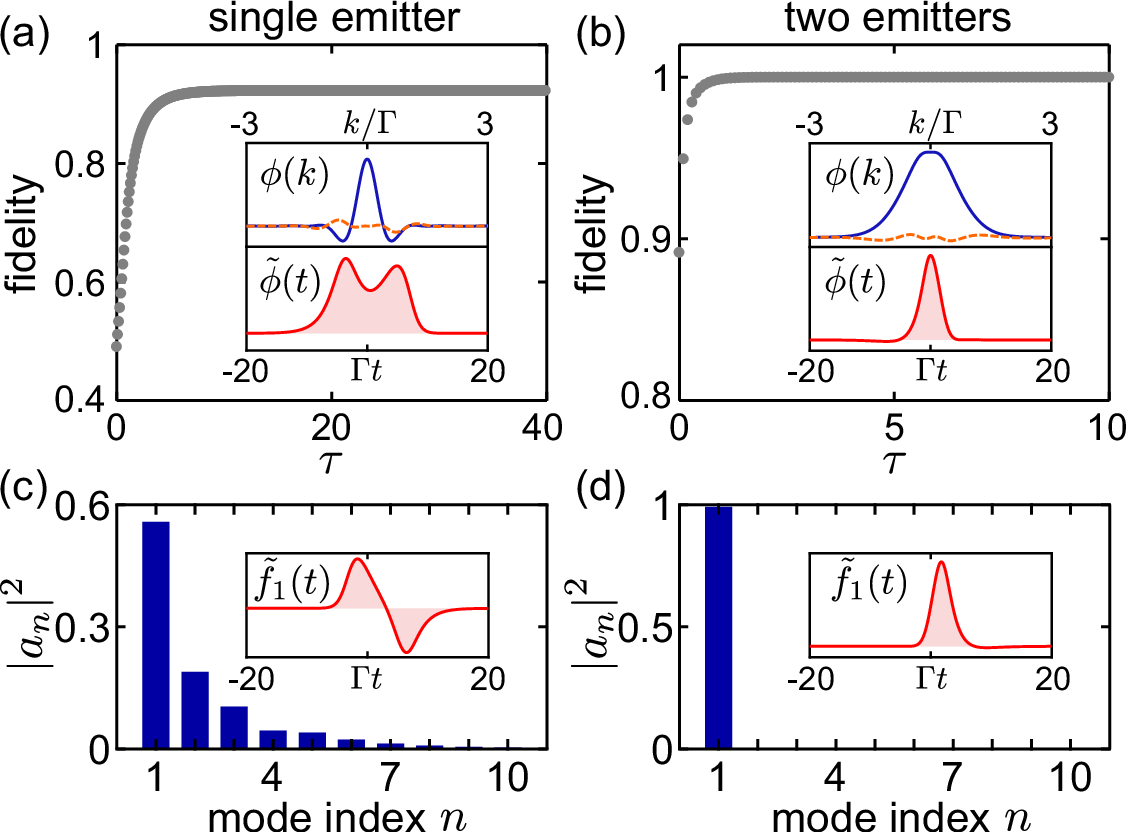}
		\caption{(a) and (b) show the sorting fidelity for a single emitter and two identical emitters during the time evolution. A Gaussian pulse $\phi(k)\propto e^{-2(k/\Gamma)^2}$ is chosen as the initial wavefunction for the time evolution. The optimal input modes are plotted in the insets in wavenumber and time domains. (c) shows the eigenvalues of the Takagi decomposition of the two-photon output state for the optimal input mode. The insets in (c) and (d) show time-domain wavefunctions of the most populated output modes. }
		\label{fig:fig3}
	\end{figure}
	
	We first apply the above optimization scheme to a single-emitter system. As shown in Fig.~\ref{fig:fig3}(a), the sorting fidelity grows monotonically during the time evolution and converges gradually to a maximum. To avoid trapping of the solution in a local optimum, we have used different types of initial trial wavefunctions, which all converge to the same optimal fidelity $\mathcal{F}\gtrsim 0.9223$. The corresponding optimal input pulse $\phi(k)$ presented in the upper inset of Fig.~\ref{fig:fig3}(a) is a complex function that possesses a major peak at $k=0$ and two minor peaks around $k=\pm~1/ \Gamma$. The time-reversal invariance of the transmission coefficient $\mathcal{T}$ and scattering matrix $\mathcal{S}$ \cite{symmetry} implies the symmetry relation $E[\phi(k),\phi^*(k)]=E[\phi^*(-k),\phi(-k)]$, and if the minimum of $E[\phi(k),\phi^*(k)]$   is nondegenerate the optimal mode must satisfy $\phi^*(-k)=\phi(k)$. This explains the  symmetry of $\phi(k)$ in momentum space and dictates that its time-domain counterpart $\tilde{\phi}(t)=\frac{1}{\sqrt{2\pi}}\int dk \phi(k)e^{-ikt}$ is a real function, as plotted in the lower inset of Fig.~\ref{fig:fig3}(a) (tilde $\sim$ is used to distinguish the time-domain wavefunction from the $k$-space one throughout this Letter).
	
	As the optimal sorting by a single emitter is far from being deterministic, we then investigate whether an improvement can be made by adding more scatterers to the system. Figure \ref{fig:fig3}(b) shows the sorting performance for two identical emitters. Surprisingly, the optimal sorting fidelity approaches unity ($\mathcal{F}\gtrsim 0.9997$) in this case, and the optimal input pulse $\tilde{\phi}(t)$ is more regular in shape, resembling a Gaussian with a slight asymmetry. To gain more insight in this significant improvement, we analyze the two-photon output wavefunction for the optimal input pulse based on the Takagi decomposition of the state [Eq.~(\ref{eq:eq1})]. For a single two-level emitter, the output state occupies multiple basis functions [see Fig.~\ref{fig:fig3}(c)] and the most populated mode $\tilde{f}_1(t)$ [inset of Fig.~\ref{fig:fig3}(c)] has a weight $|a_1|^2\sim 0.5579$. An unexpected phenomenon occurs when two emitters are included: the output photons are confined to a single temporal mode $\tilde{f}_1(t)$ with a probability $|a_1|^2\gtrsim 0.9911$ [see Fig.~\ref{fig:fig3}(d)], whose shape approaches the time-reversed input pulse, i.e., $\tilde{f}_1(t) \approx \tilde{\phi}(t_{d}-t)$ with $t_{d}$ a delay time. We then examine the wavefunction $\Psi_m(k_1,k_2)$ of the field between the two emitters. As illustrated in Fig.~\ref{fig:fig1}(b), the two-photon state is entangled over several temporal modes due to the first scatterer. However, the optimal pulse appears to be a special input mode, which under the action of the single-emitter scattering matrix $\mathcal{S}_0$ generates an output wavefunction satisfying $\Psi_m(k_1,k_2) \approx \Psi_m(-k_1,-k_2)e^{i(k_1+k_2) t_{d}}$. This relation makes the second scattering process a shifted time reversal of the first one, i.e., $\int dp_1 dp_2\mathcal{S}_0(k_1,k_2;p_1,p_2)\Psi_m(p_1,p_2)\approx \phi(-k_1)\phi(-k_2)e^{i(k_1+k_2)t_d}$ \cite{supply}. The underlying physics of the scattering then intuitively explains the nearly perfect sorting enabled by the emitter pair
	\begin{equation}
	\alpha \hat{a}_{\phi}^\dagger|0\rangle + 
	\beta\frac{ (\hat{a}_{\phi}^\dagger)^2}{\sqrt{2}}|0\rangle
	~ \longrightarrow ~ \alpha \hat{a}_{\psi}^\dagger|0\rangle + 
	\beta\frac{(\hat{a}_{f_1}^\dagger)^2 }{\sqrt{2}}|0\rangle. \label{eq:eq8}
	\end{equation}
	First, for a single-photon input, the dispersion of the Gaussian-like pulse will accumulate instead of being cancelled through successive interactions with the emitters, which results in a distorted
	output wavefunction $\tilde{\psi}(t)$ [see Fig.~\ref{fig:fig1}(a)]. However, when two photons are injected, the above quasi-time-reversal process makes the output mode $\tilde{f}_1(t)$ almost free of distortion and it can be made orthogonal to $\tilde{\psi}(t)$ by slight adjustment of the input mode.
	
\begin{figure}
		\centering
		\includegraphics[width=\linewidth]{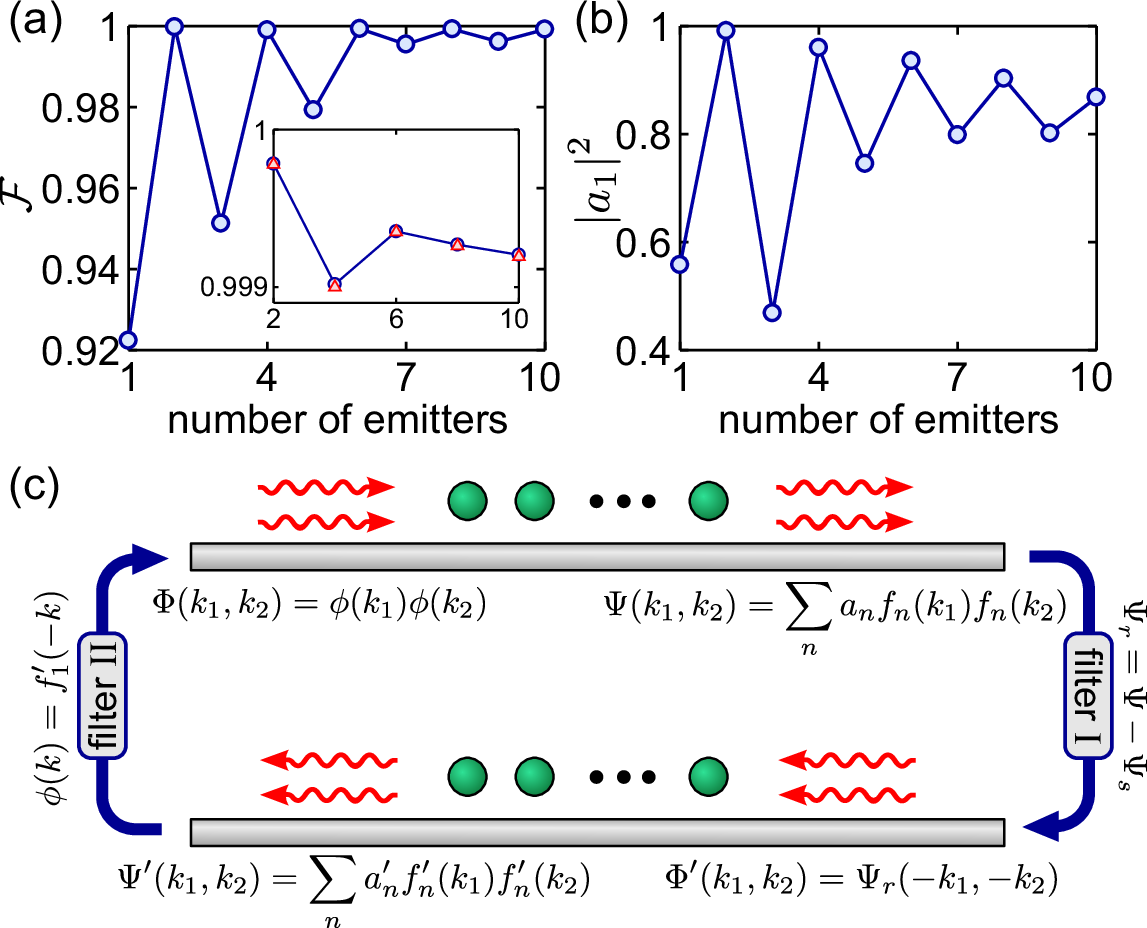}
		\caption{(a) and (b) show the optimal fidelity $\mathcal{F}$ and  the most populated mode probability $|a_1|^2$ as a function of the number of emitters. The blue dots are obtained by the time-marching method. The red triangles are obtained with the iterative filtering protocol shown schematically in (c).}
		\label{fig:fig4}
	\end{figure}

	We have employed the generality of the above time-reversal sorting principle and performed the optimization for systems containing $N_e$ identical emitters. As shown in Fig.~\ref{fig:fig4}(a), the optimal sorting process strongly depends on the parity of $N_e$. For odd $N_e$, increasing the emitter number can gradually improve the sorting performance. When $N_e$ is even, the second half of the system can induce an effective time reversal of the two-photon scattering induced by the first half, leading to generally high sorting fidelities ($\mathcal{F}>0.999$). However, larger numbers of emitters do not exceed the already high fidelity of $N_e=2$ [see Fig.~\ref{fig:fig4}(b)].
	
	In addition to the time-marching gradient method, we also develop an iterative optimization scheme [see Fig.~\ref{fig:fig4}(c)], which only requires the output state instead of the full knowledge of the scattering matrix and thus makes it more suitable for experimental implementation. The iteration is composed of a forward scattering and its time reversal process interspersed by two filters, where the first one filters out the undesired wavefunction $\Psi_s$, and the second filter extracts the time-reversed most populated mode $f_1^\prime(-k)$ from the output wavefunction $\Psi^\prime(k_1,k_2)$. Iteration by this filter is reminiscent of the one used in optimal quantum storage \cite{gorshkov2007universal,gorshkov2007photon,novikova2007optimal}, except that our state mapping $\phi_{[n+1]}=\hat{M}\phi_{[n]}$ is nonlinear. It is straightforward to verify that a perfect sorting should be a fixed point of the iteration, but a rigorous proof of convergence to the optimum is difficult as the successive nonlinear mapping $\hat{M}$ cannot be interpreted as a power iteration as in the linear storage problem \cite{gorshkov2007universal}. Nonetheless, for even values of $N_e$ we have always found convergence and an equally high fidelity as by the time-marching method [see inset of Fig.~\ref{fig:fig4}(a)].
	
	{\it Experimental considerations}.---Since incorporating more emitters to the system is experimentally challenging and does not improve the fidelity for $N_e>2$, we recommend use of the emitter-pair based photon sorter. To assess if our scheme is feasible for state-of-the-art implementations of the waveguide QED platform, we now consider some realistic imperfections. 
	
\begin{figure}
		\centering
		\includegraphics[width=\linewidth]{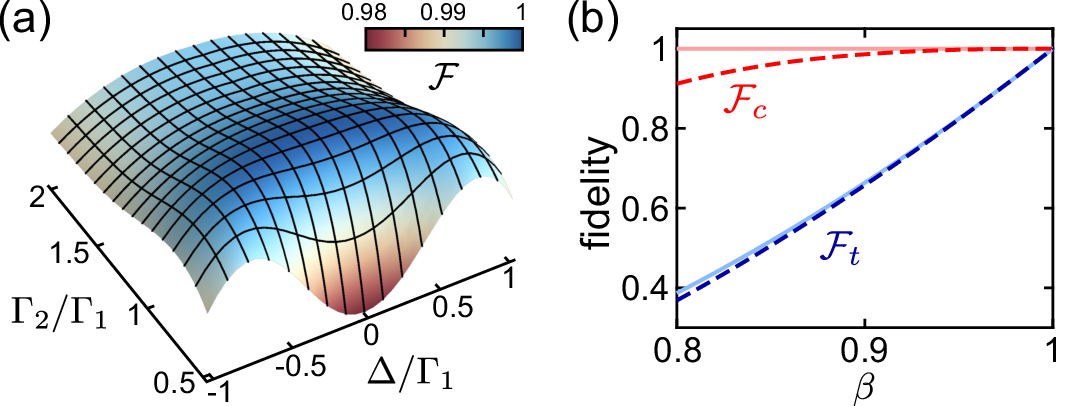}
		\caption{(a) Optimal fidelity $\mathcal{F}$ as a function of $\Gamma_2/\Gamma_1$ and $\Delta/\Gamma_1$ for two nonidentical emitters. (b) Total fidelity $\mathcal{F}_t$ (blue lines) and conditional fidelity $\mathcal{F}_c$ (red lines) for finite directional efficiency factors $\beta$. The solid lines are obtained by minimizing $E-N_2$ and $E/N_2$ at each $\beta$, while the dashed lines are obtained by using input mode that have been optimized for $\beta=1$.}
		\label{fig:fig5}
	\end{figure}

	First, we investigate the sensitivity of our scheme to deviations between the properties of the two emitters. In particular, we consider a difference in the emitter-photon coupling strengths $\Gamma_1$, $\Gamma_2$ and a detuning $\Delta=\nu_2-\nu_1$ between their resonance frequencies. To that end, we perform the input mode optimization for a range of different values of  $\Gamma_2/\Gamma_1$ and $\Delta/\Gamma_1$. As shown in Fig.~\ref{fig:fig5}(a), the optimal fidelity $\mathcal{F}$ remains high for a broad range of parameters assuring that fabrication issues may not be an impediment to the scheme. We do find that the order of emitters plays a role, and for a small imbalance $\Gamma_2/\Gamma_1\lesssim 1$, the optimal fidelity can even be higher than in the case of identical emitters, e.g., $\mathcal{F}>0.9999$ can be obtained at $\Delta=0$, $\Gamma_2/\Gamma_1=0.95$. When the coupling strengths are largely imbalanced,  $\Gamma_2>\Gamma_1$ is preferable near $\Delta=0$. The order of emitters is not important when the coupling strengths are identical, because the optimal fidelity is a symmetric function of the detunings [$\mathcal{F}(\Delta)=\mathcal{F}(-\Delta)$].
	
	Second, we address the case of imperfect emitter-photon couplings, for which photon losses into free space are assumed and the directional efficiency factor $\beta=\Gamma/\Gamma_\mathrm{tot}$ is introduced to quantify the fraction of the total emitter decay that leads to light emission in the desired waveguide mode. When this quantity is less than unity, the single- and two-photon output wavefunctions $\psi(k)$ and $\Psi(k_1,k_2)$ become unnormalized, and we need to calculate their respective survival probabilities $N_1$ and $N_2$
	\begin{equation}
	N_1 = \int dk |\psi(k)|^2,\quad
	N_2 = \int dk_1 dk_2 |\Psi(k_1,k_2)|^2. \label{eq:eq9}
	\end{equation}
	As a result, the Hermitian kernel in Eq.~(\ref{eq:eq6}) is modified into the following form
	\begin{equation}
	\mathcal{H}= 2N_1^{-1}\int dk \mathcal{L}_1^*(k,p)\mathcal{L}_1(k,p^\prime)-N_1^{-2}\mathcal{L}_2^*(p) \mathcal{L}_2(p^\prime). \label{eq:eq10}
	\end{equation}
	The fidelity to be optimized can be defined in different ways, e.g., the total fidelity $\mathcal{F}_t=N_2-E$ gives the norm of the desired two-photon component $\Psi_r(k_1,k_2)$, and the conditional fidelity $\mathcal{F}_c=\mathcal{F}_t/N_2=1-E/N_2$ characterizes the sorting performance conditioned on survival of both photons. To maximize $\mathcal{F}_t$ and $\mathcal{F}_c$, we must minimize $E(\phi,\phi^*)-N_2(\phi,\phi^*)$ and $E(\phi,\phi^*)/N_2(\phi,\phi^*)$, respectively. As shown in Fig.~\ref{fig:fig5}(b), the fidelity optimized for a given $\beta$ based on the above modified scheme outperforms the one merely using the same mode as was optimized for $\beta=1$. We also note that while the photon loss causes a proportional reduction in $\mathcal{F}_t$, a very high conditional fidelity $\mathcal{F}_c> 0.9997$ can be maintained for $\beta<1$, and hence a heralded photon sorter can be constructed with lossy emitters.
	
	In conclusion, we have presented a theoretical analysis of the sorting of Fock states into orthogonal modes by their coupling to two-level emitters. For the chiral waveguide QED system considered in this Letter, a near-unity sorting of one- and two-photon states can be achieved by two or a higher even numbers of emitters. The proposed optimal photon sorter, combined with single-qubit operations (e.g., temporal mode extraction \cite{eckstein2011quantum,ansari2018tomography,ansari2018tailoring} and pulse time reversal \cite{chumak2010all,sivan2011time,minkov2018localization}) can be used to build advanced quantum photonic devices \cite{ralph2015photon,pick2021boosting}, such as deterministic Bell state analyzers and nonlinear-sign gates \cite{knill2001sche} with fidelity $\gtrsim 0.9995$ \cite{supply}. It is also possible to optimize the scheme against degradation of the fidelity due to pure dephasing of the emitter \cite{supply}. By applying multilevel \cite{iversen2021strongly,iversen2021self} or driven \cite{fischer2018scattering} systems, it may be possible to further improve the sorting performance, and generalization to the multiphoton regime for high-dimensional sorting \cite{mahmoodian2020dynamics} presents an attractive topic of further exploration. 

	\begin{acknowledgments}
		We acknowledge valuable discussions with T. C. Ralph, Ole A. Iversen, Lida Zhang, and Jesper Hasseriis Mohr Jensen. This work is supported by the Carlsberg Foundation through the ``Semper Ardens'' Research Project QCooL, and by the Danish National Research Foundation (Centers of Excellence CCQ DNRF156 and Hy-Q DNRF139).
	\end{acknowledgments}
	
	\bibliography{main_text}

\begin{thebibliography}{42}
\bibitem [{sym()}]{symmetry}%
  \BibitemOpen
  \href@noop {} {}\bibinfo {note} {The explicit relations are given by
  $\mathcal{T}(-k)=\mathcal{T}^*(k)$ and
  $\mathcal{S}(-k_1,-k_2;-p_1,-p_2)=\mathcal{S}^*(k_1,k_2;p_1,p_2)$.}\BibitemShut
  {Stop}%
\end{thebibliography}
	
\end{document}


\preprint{APS/123-QED}
	
	\title{Supplementary Material for ``Deterministic Photon Sorting in Waveguide QED Systems''}
	
	\author{Fan Yang}
	\affiliation{Center for Complex Quantum Systems, Department of Physics and Astronomy, Aarhus University, DK-8000 Aarhus C, Denmark}
	
	\author{Mads M. Lund}
	\affiliation{Center for Complex Quantum Systems, Department of Physics and Astronomy, Aarhus University, DK-8000 Aarhus C, Denmark}
	
	\author{Thomas Pohl}
	\affiliation{Center for Complex Quantum Systems, Department of Physics and Astronomy, Aarhus University, DK-8000 Aarhus C, Denmark}
	
	\author{Peter Lodahl}
	\affiliation{Center for Hybrid Quantum Networks (Hy-Q), Niels Bohr Institute, University of Copenhagen, Blegdamsvej 17, DK-2100 Copenhagen, Denmark}
	
	
	\author{Klaus M{\o}lmer}
	\affiliation{Center for Complex Quantum Systems, Department of Physics and Astronomy, Aarhus University, DK-8000 Aarhus C, Denmark}
	\affiliation{Aarhus Institute of Advanced Studies, Aarhus University, DK-8000 Aarhus C, Denmark}
	
	\maketitle
	\onecolumngrid
	
	This supplementary note provides several technical details of the main text, including: (i) exact form of the gradient of the error functional (Sec.~\ref{sec:sec1}); (ii) discussion of the time-reversal scattering (Sec.~\ref{sec:sec2}); (iii) application of the photon sorter in a Bell-state analyzer and photonic nonlinear-sign gate (Sec.~\ref{sec:sec3}); (iv) influence of the pure dephasing (Sec.~\ref{sec:sec4}).
	
	\section{Gradient of the Energy functional}\label{sec:sec1}
	In the main text, we derive an explicit form of the error functional $E(\phi,\phi^*) = \int dp dp^\prime \phi^*(p) \mathcal{H}(p,p^\prime) \phi(p^\prime)$, where the Hermitian kernel $\mathcal{H}(p,p^\prime)$ is given by
	\begin{equation}
	\mathcal{H}(p,p^\prime) = 2\int dk \mathcal{L}_1^*(k,p)\mathcal{L}_1(k,p^\prime)-\mathcal{L}_2^*(p) \mathcal{L}_2(p^\prime), \label{eq:eq1}
	\end{equation}
	with $\mathcal{L}_1(k,p)$ and $\mathcal{L}_2(p)$ defined as
	\begin{equation}
	\mathcal{L}_1(k,p) =\int dk_1dp_1 \mathcal{S}(k_1,k;p_1,p)\mathcal{T}^*(k_1)\phi^*(k_1)\phi(p_1), \quad
	\mathcal{L}_2(p) =\int dk \mathcal{L}_1(k,p) \mathcal{T}^*(k)\phi^*(k). \label{eq:eq2}
	\end{equation}
	The optimization is based on the time-marching gradient method
	\begin{equation}
	\partial_\tau \phi = -\frac{\delta E(\phi,\phi^*)}{\delta \phi^*}. \label{eq:eq3}
	\end{equation}
	The gradient of $E(\phi,\phi^*)$ with respect to $\phi^*$ is determined by ${\delta E}/{\delta \phi^*} = \int dp^\prime \mathcal{M}(p,p^\prime)\phi(p^\prime)$, where $\mathcal{M}(p,p^\prime)$ reads
	\begin{align}
	\mathcal{M}(p,p^\prime) = 4\int dk~ \mathcal{L}^*_1(k,p)\mathcal{L}_1(k,p^\prime) 
	-2\mathcal{L}^*_2(p)\mathcal{L}_2(p^\prime) +2\mathcal{L}_3(p,p^\prime)-2\mathcal{C}t^*(p)\mathcal{L}_1(p,p^\prime),
	\end{align}
	with $\mathcal{L}_3(p,p^\prime)$ and $\mathcal{C}$ defined as
	\begin{align}
	\mathcal{L}_3(p,p^\prime) = \int dk_1 dp_1  ~ \mathcal{S}(k_1,p;p_1,p^\prime)f(p_1)\int dk \mathcal{L}_1^*(k_1,k)f^*(k),\quad
	\mathcal{C} = \int dk~f^*(k)\mathcal{L}_2^*(k).
	\end{align}
	
	\section{Self-time-reversal two-photon scatterings}\label{sec:sec2}
	
\begin{figure}
		\centering
		\includegraphics[width=\linewidth]{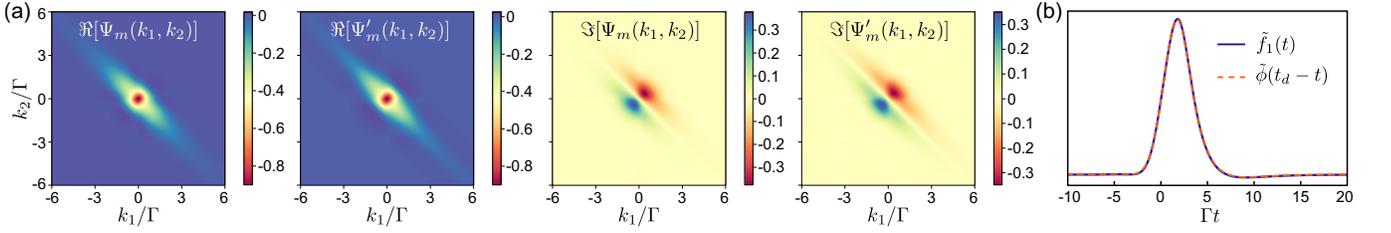}
		\caption{(a) Real and imaginary part of the two-photon wavefunction $\Psi_m(k_1,k_2)$ and $\Psi_m^\prime(k_1,k_2)$. (b) Comparison between the most populated output mode $\tilde{f}_1(t)$ and the shifted time-reversed input mode $\tilde{\phi}(t_d-t)$.}
		\label{fig:fig1}
	\end{figure}

	We find numerically that while a single emitter splits any single-mode two-photon pulse into several modes, for an optimally chosen input mode the multimode field state returns to a single mode by the scattering on a second emitter. We can ascribe this result to an approximate self-time-reversal of the two-photon scattering events, i.e., the output two-photon wavefunction $\Psi_m(k_1, k_2)$ of the first scattering process evolves into a delayed and time reversed copy of the input state by the second scattering process. For this to happen, the intermediate wavefunction $\Psi_m(k_1, k_2)$ must satisfy $\Psi_m(k_1,k_2) \approx \Psi_m(-k_1,-k_2)e^{i(k_1+k_2) t_{d}}$ with the time delay $t_d$. In Fig.~\ref{fig:fig1}(a), we check by direct comparison of $\Psi_m(k_1, k_2)$ and $\Psi_m^\prime(k_1,k_2)=\Psi_m(-k_1,-k_2)e^{i(k_1+k_2) t_{d}}$ that this property is, indeed, well fulfilled with the delay parameter $t_d \approx 1.81/\Gamma$. This explains why the output photons populate a single temporal mode $\tilde{f}_1(t)$ close to $\tilde{\phi}(t_d-t)$ with very high fidelity [see Fig.~\ref{fig:fig1}(b)]. 
	
	\section{Efficient Bell-state Measurement and Nonlinear Sign Gate}\label{sec:sec3}
\begin{figure}
		\centering
		\includegraphics[width=\linewidth]{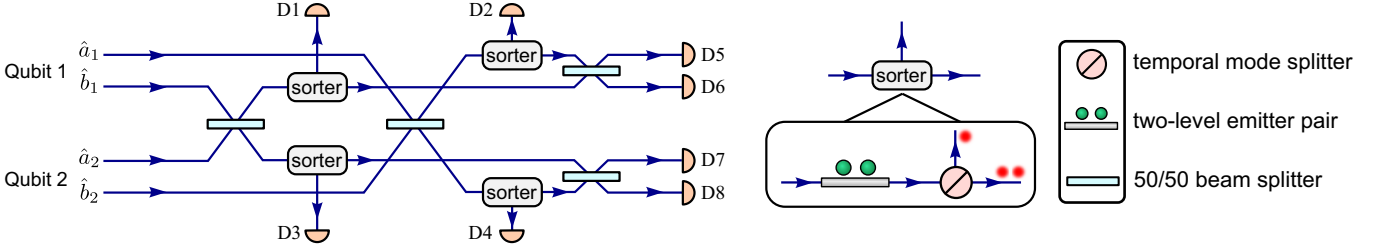}
		\caption{Setup for photon-sorter based Bell-state analyzer. The temporal mode beam splitting can be efficiently performed with the sum-frequency generation and a dichroic beam splitter \cite{eckstein2011quantum,ansari2018tomography,ansari2018tailoring}.}
		\label{fig:fig2}
	\end{figure}

	In this section, we present two realistic applications of the emitter-pair based photon sorter discussed in the main text. We first discuss how to use the near-deterministic photon sorter to perform an efficient Bell-state measurement \cite{ralph2015photon}. The setup is shown in Fig.~\ref{fig:fig2}, where input photons are encoded in qubit states $|a\rangle=\hat{a}^\dagger|0\rangle$ and $|b\rangle=\hat{b}^\dagger|0\rangle$. Now let us analyze the detection outcomes for different input Bell states. For the Bell-state $ |\Phi_\pm\rangle=(|a_1a_2\rangle \pm |b_1b_2\rangle)/\sqrt{2}$, there are only single-photon components entering each photon sorter, and the measurement outcomes are completely distinguishable as summarized in Table~\ref{tab:table1}. For the Bell-state $ |\Psi_\pm\rangle=(|a_1b_2\rangle \pm |b_1a_2\rangle)/\sqrt{2}$, the sorter has a small probability $1-\mathcal{F}$ to fail, but the failure can be faithfully heralded by false clicks, including double clicks of a detector from D1-D4, and mixed clicks of a detector within D1-D4 and another detector within D5-D8. The successful clicks are summarized in Table~\ref{tab:table1}, which have a near-unity success probability $\mathcal{F}\gtrsim 0.9997$.

	\begin{table}
		\caption{\label{tab:table1}%
			Indications of successful events and their corresponding probabilities for different input Bell states.
		}
		\begin{ruledtabular}
			\begin{tabular}{lcc}
				\multicolumn{1}{c}{\textrm{input Bell state}}& 
				\multicolumn{1}{c}{\textrm{successful clicks}}&
				\multicolumn{1}{c}{\textrm{success probability}}\\ 
				\hline
				$|\Phi_+\rangle=(|a_1a_2\rangle+|b_1b_2\rangle)/\sqrt{2}$   & \{D1,D2\} or \{D3,D4\}& 1 \\
				$|\Phi_-\rangle=(|a_1a_2\rangle-|b_1b_2\rangle)/\sqrt{2}$   & \{D1,D4\} or \{D2,D3\}& 1 \\			$|\Psi_+\rangle=(|a_1b_2\rangle+|b_1a_2\rangle)/\sqrt{2}$   & \{D5,D6\} or \{D7,D8\}& $\mathcal{F}\gtrsim 0.9997$ \\			$|\Psi_-\rangle=(|a_1b_2\rangle-|b_1a_2\rangle)/\sqrt{2}$   & \{D5\} or \{D6\} or \{D7\} or \{D8\}& $\mathcal{F}\gtrsim 0.9997$ \\
			\end{tabular}
		\end{ruledtabular}
	\end{table}
	
\begin{figure}
		\centering
		\includegraphics[width=0.66\linewidth]{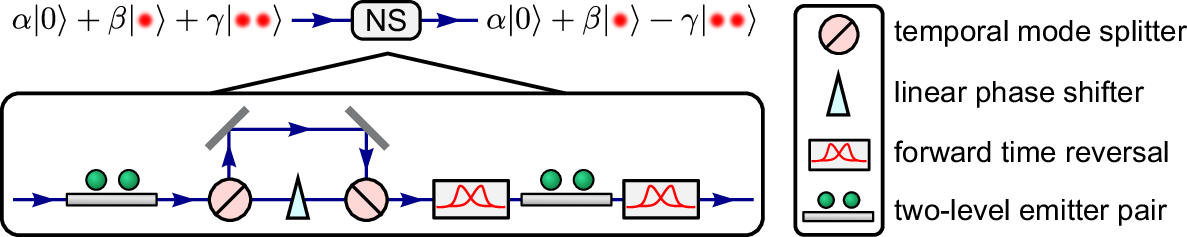}
		\caption{Schematic of the optical configuration for realizing a nonlinear-sign (NS) gate with the proposed photon sorter.}
		\label{fig:fig3}
	\end{figure}

	Next, we discuss the implementation of a photonic nonlinear-sign (NS) gate \cite{witthaut2012photon,ralph2015photon}, which is a key element in logic-gate based all-optical quantum computing \cite{knill2001sche}. An ideal NS gate transforms the input state as
	\begin{equation}
	\alpha |0\rangle + \beta \hat{a}_{\phi}^\dagger|0\rangle + 
	\gamma\frac{ (\hat{a}_{\phi}^\dagger)^2}{\sqrt{2}}|0\rangle
	~ \longrightarrow ~  \alpha |0\rangle + \beta \hat{a}_{\phi}^\dagger|0\rangle - 
	\gamma\frac{ (\hat{a}_{\phi}^\dagger)^2}{\sqrt{2}}|0\rangle,
	\end{equation}
	i.e., it only flips the sign of the two-photon component, and leaves the vacuum and single-photon part unchanged. Such a nonlinear operation can be realized by the established photon sorter and several linear-optic elements, as shown in Fig.~\ref{fig:fig3}. The idea is to use a pair of photon sorters to build a Mach-Zehnder-like interferometer, with which a linear phase flip can be applied only to the two-photon component. Two time reversal operations of the pulse \cite{chumak2010all,sivan2011time,minkov2018localization} are also needed to convert the multimode output state to its initial input mode $\phi(k)$.
	
	To analyze the fidelity of the NS gate, we consider a two-photon input state, i.e., $\alpha=\beta=0$, $\gamma=1$. In this case, after passing through the first emitter-pair based photon sorter, the two-photon wavefunction $\Psi_0(k_1,k_2) = \int dp_1dp_2 \mathcal{S}(k_1,k_2,p_1,p_2)\phi(p_1)\phi(p_2)$ can be decomposed into the following form (see the main text)
	\begin{equation}
	\Psi_0(k_1,k_2) = c_2\psi(k_1)\psi(k_2)+\frac{c_1}{\sqrt{2}}[\psi(k_1)\theta(k_2)+\theta(k_1)\psi(k_2)]+\Psi_r(k_1,k_2),
	\end{equation}
	where $\Psi_r(k_1,k_2)=\sum_{n}a_n \mu_n^2 {\theta}_n(k_1) {\theta}_n(k_2)$ is the desired output wavefunction. Transmitting through the linear phase shifter that contributes an extra $\pi/2$ phase to all $\theta_n(k)$ modes, and combined at the second temporal mode beam splitter, the wavefunction is modified into the following form,
	\begin{equation}
	\Psi_1(k_1,k_2) = c_2\psi(k_1)\psi(k_2)+i\frac{c_1}{\sqrt{2}}[\psi(k_1)\theta(k_2)+\theta(k_1)\psi(k_2)]-\Psi_r(k_1,k_2).
	\end{equation}
	A pair of time reversal operations together with the second emitter pair finally transform the two-photon state into
	\begin{equation}
	\Psi_2(k_1,k_2) = \int dp_1dp_2 \mathcal{S}(-k_1,-k_2,p_1,p_2)\Psi_1(-p_1,-p_2). \label{eq:eq9}
	\end{equation}
	The fidelity $\mathcal{F}_\mathrm{NS}$ and the induced phase shift $\varphi_\mathrm{NS}$ of the NS gate operation can be determined by
	\begin{equation}
	\sqrt{\mathcal{F}_\mathrm{NS}} ~ e^{i\varphi_\mathrm{NS}} = \int dk_1dk_2 [\phi(k_1)\phi(k_2)]^* \Psi_2(k_1,k_2). \label{eq:eq10}
	\end{equation}
	Substituting Eq.~(\ref{eq:eq9}) into Eq.~(\ref{eq:eq10}), and using the unitary property of the scattering matrix ($\mathcal{S}^\dagger \mathcal{S}=1$), we obtain
	\begin{equation}
	\sqrt{\mathcal{F}_\mathrm{NS}} ~ e^{i\varphi_\mathrm{NS}} = \int dk_1dk_2 \Psi_0^*(k_1,k_2)\Psi_1(k_1,k_2) = 2|c_2|^2+\sqrt{2}e^{i\pi/4}|c_1|^2-1.
	\end{equation}
	It is straightforward to check that a deterministic sorter ($c_1=c_2=0$) can give an ideal NS gate operation with $\mathcal{F}_\mathrm{NS}=1$ and $\varphi_\mathrm{NS}=\pi$. For the emitter-pair based photon sorter studied in the main text, the unwanted populations can be suppressed to $|c_1|^2\approx 1.8\times 10^{-4}$ and $|c_2|^2\approx 3.4\times 10^{-5}$ when the optimal input mode is used, which ultimately gives a near-deterministic NS gate operation with $\mathcal{F}_\mathrm{NS}\gtrsim 0.9995$ and $\varphi_\mathrm{NS}/\pi \approx 0.99994$.
	
	\section{Influence of Pure Dephasing}\label{sec:sec4}
\begin{figure}
		\centering
		\includegraphics[width=0.72\linewidth]{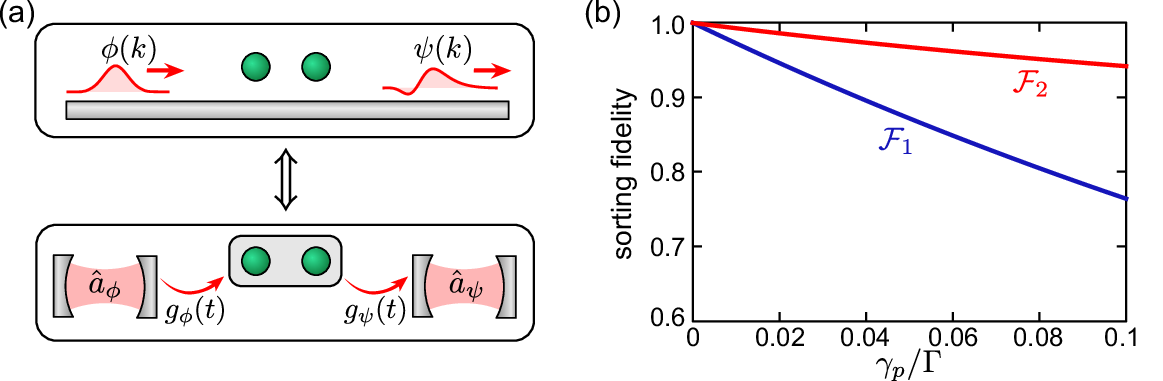}
		\caption{(a) Illustration of the input-output quantum pulse method to determine the output photonic state of a given temporal mode $\tilde{\psi}(t)$. (b) Sorting fidelity as a function of the dephasing rate $\gamma_p$.}
		\label{fig:fig4}
	\end{figure}

	When taking pure dephasing of the emitter into account, the impure output photons cannot be described by wavefunctions. In this case, the dynamics are conveniently treated by the input-output quantum pulse method \cite{kiilerich2019input}. With this method [see Fig.~\ref{fig:fig4}(a)], the scattering into the output mode $\psi(k)$ from an input mode $\phi(k)$ can be determined by considering an imaginary cascaded system containing upstream and downstream cavity modes $\hat{a}_\phi$ and $\hat{a}_\psi$ with time-dependent couplings $g_\phi(t)$ and $g_\psi(t)$ to the waveguide QED system, whose exact forms are given by
	\begin{equation}
	g_\phi(t) = \frac{\tilde{\phi}^*(t)}{\sqrt{1-\int_0^t dt^\prime |\tilde{\phi}(t^\prime)|^2}}, \qquad 
	g_\psi(t) = -\frac{\tilde{\psi}^*(t)}{\sqrt{\int_0^t dt^\prime |\tilde{\psi}(t^\prime)|^2}}.
	\end{equation}
	Then, the dynamics of such a cascaded system can be described by a master equation $\partial_t\hat{\rho}=-i[\hat{H},\hat{\rho}]+\sum_i\mathcal{D}[\hat{L}_i]\hat{\rho}$, where the Hamiltonian and the Lindblad operators are given by
	\begin{align}
	\hat{H} &= \frac{i}{2}\left[\sqrt{\Gamma}g_\phi(t)\hat{a}_\phi^\dagger (\hat{\sigma}_{ge}^{(1)}+\hat{\sigma}_{ge}^{(2)}) + \sqrt{\Gamma}g^*_\psi(t)(\hat{\sigma}_{eg}^{(1)}+\hat{\sigma}_{eg}^{(2)}) \hat{a}_\psi + g_\phi(t)g^*_\psi(t)\hat{a}_\phi^\dagger\hat{a}_\psi
	+ \Gamma \hat{\sigma}_{eg}^{(1)}\hat{\sigma}_{ge}^{(2)} -\mathrm{H.c.} \right],\\
	\hat{L}_0 &= \sqrt{\Gamma}(\hat{\sigma}_{ge}^{(1)}+\hat{\sigma}_{ge}^{(2)}) + g_\phi^*(t)\hat{a}_\phi + g_\psi^*(t)\hat{a}_\psi,\qquad \hat{L}_1 ={\sqrt{\gamma_p}\hat{\sigma}_{ee}^{(1)}}, \qquad \hat{L}_2 ={\sqrt{\gamma_p}\hat{\sigma}_{ee}^{(2)}},
	\end{align}
	with $\hat{\sigma}^{(i)}_{\alpha\beta}=|\alpha\rangle_i \langle \beta|$ the spin operator for the $i$-th emitter, and $\gamma_p$ the dephasing rate between the ground state $|g\rangle_i$ and the excited state $|e\rangle_i$ of the emitter.
	
	The performance of the photon sorting can be characterized by the reduced density matrix of the downstream cavity $\hat{a}_\psi$. For the single-photon and the two-photon input, the reduced density matrices are respectively given by
	\begin{equation}
	\hat{\rho}^{(1)}_\psi=\begin{pmatrix}
	\rho_{00}^{(1)} & 0\\
	0 & \rho_{11}^{(1)}
	\end{pmatrix} \quad \textrm{and} \quad
	\hat{\rho}^{(2)}_\psi=\begin{pmatrix}
	\rho_{00}^{(2)} & 0 & 0\\
	0 & \rho_{11}^{(2)} & 0\\
	0 & 0 & \rho_{22}^{(2)}
	\end{pmatrix}.
	\end{equation}
	Here, we need to introduce two different fidelities, where $\mathcal{F}_1=\rho_{11}^{(1)}$ represents the fidelity of directing a single photon into the desired mode $\psi(k)$, and  $\mathcal{F}_2=\rho_{00}^{(2)}$ denotes the fidelity of sorting the two-photon state into modes (while not necessarily a single mode) orthogonal to $\psi(k)$. A deterministic photon sorter requires $\mathcal{F}_1=\mathcal{F}_2=1$, and the emitter pair of the main text can give a near-deterministic performance with $\mathcal{F}_1=1$ and $\mathcal{F}_2\gtrsim 0.9997$. In the presence of pure dephasing ($\gamma_p\neq 0$), both $\mathcal{F}_1$ and $\mathcal{F}_2$ deviate from unity, and $\mathcal{F}_1$ falls off faster than $\mathcal{F}_2$ as the dephasing rate increases [see Fig.~\ref{fig:fig4}(b)]. When proper temporal mode filtering is applied, the influence induced by dephasing can be mitigated by constructing a heralded photon sorter as for the case of imperfect coupling discussed in the main text.
	
	\bibliography{supplement}